# Achieving Cooling Without Repump Lasers Through Ion Motional Heating

*Yue Xiao[1,2], Yongxu Peng[1,2], Linfeng Chen[1,2], Chunhui Li[1], Zongao Song[1], Xin Wang[1], Tao Wang[1], Yurun Xie[1], Bin Zhao[1]\*, Tiangang Yang[1]\**

1. Department of Chemistry, and Center for Advanced Light Source, Southern University of Science and Technology, Shenzhen, Guangdong 518055, China.
2. These authors contributed equally.

\* Corresponding Author: binz@sustech.edu.cn, yangtg@sustech.edu.cn

**ABSTRACT:** Laser cooling typically requires one or more repump lasers to clear dark states and enable recycling transitions. Here, we have achieved cooling of $Be^+$ ions using a single laser beam, facilitated by one-dimensional heating through micromotion. By manipulating the displacement from the trap's nodal line, we precisely controlled the ion micromotion direction and speed, reaching up to 3144 m/s, which corresponds to a 7.1 GHz Doppler frequency shift in our experiment. This approach eliminates the necessity of a 1.25 GHz offset repump laser while keeping the $Be^+$ ions cold in the perpendicular direction. Measurements were taken using cooling laser detuning and imaging of ion trajectories. Molecular dynamics simulations, based on machine learned time-dependent electric field $E(X, Y, Z, t)$ inside the trap, accurately reproduced the experimental observation, illuminating the relationship between the direction of micromotion and the trapping electric filed vector. This work not only provides a robust method for managing the micromotion velocity of ions but also sheds light on laser cooling complex systems that require multiple repumping lasers. Additionally, it offers a method for controlling energy in the context of ion-molecule collision investigations.

TOC GRAPHICS

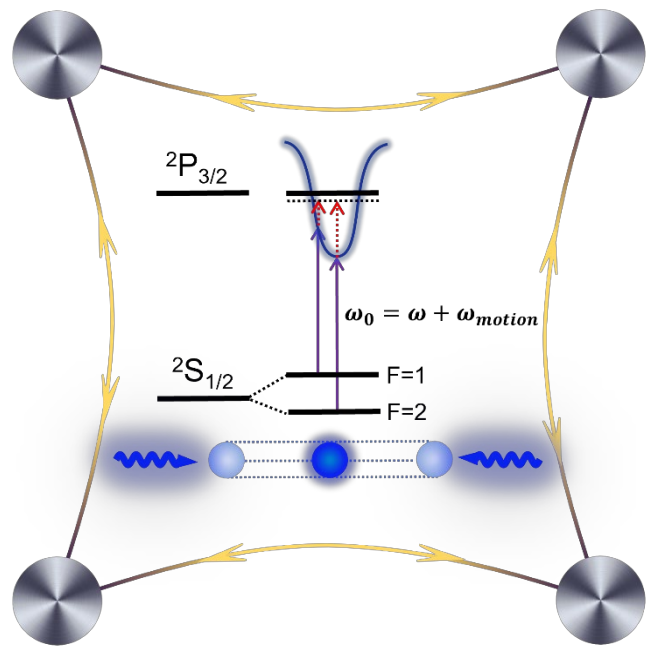

**INTRODUCTION**

In ion trap physics, micromotion arises from ions responding to the periodic driving forces of radio-frequency (RF) fields [1-3]. At the geometric center of the trap, known as the nodal line, the electric field intensity is zero, minimizing ion micromotion. However, as ions are displaced from the nodal line, the amplitude of their micromotion increases, a phenomenon referred to as excess micromotion [4-7]. Precise control of micromotion is crucial in quantum information processing (QIP) [8-12] and precision measurement experiments [13,14]. Techniques such as micromotion compensation

[4,5,15], pulsed cooling [6], artificial neural networks [16], Ramsey interferometry [17], cross-correlation [2,7,18] and resolved sideband [19-21] are employed to measure and reduce the micromotion.

In specific experiments, enhancing the micromotional states of ions proves to be beneficial. For instance, in QIP, certain non-adiabatic entanglement gates exploit ion micromotion to boost quantum entanglement fidelity [22], or to enable precise laser targeting of individual ions [23]. Moreover, research utilizing RF-driven micromotion ions as self-sustained oscillators [24-26], as well as in precision measurements, such as calculating the second-order Doppler frequency shift for atomic clocks based on large ion ensembles, necessitates strict control over micromotion velocity throughout the entire ion trap [27,28]. In studies of chemical collision processes, managing ion velocities within the trap to adjust collision energies provides valuable insights into chemical reaction dynamics [29-32]. Consequently, the impact of micromotion on experimental outcomes varies according to the objectives of the study, necessitating optimized control based on specific requirements.

The concept of "cooling by heating" involves increasing the occupation of a mode that couples the system to a cold bath, suggesting the potential use of incoherent thermal light sources or unfiltered sunlight to cool a quantum system [33-35]. Broadband lasers or femtosecond lasers that produce a wide spectrum of light are employed to address multiple energy level transitions, achieving broadband vibrational or rotational optical cooling [36-38]. These approaches could aid in the laser cooling of complex systems, such as molecular ions, where the intricate vibrational and rotational energy level structures make direct laser cooling challenging [39,40].

Here, we propose cooling Be⁺ ions by utilizing one-dimensional micromotion heating of ions. The velocity of ion oscillations facilitates a scanned laser detuning due to the Doppler effect, allowing a single laser beam to cover multiple transition energy levels, thereby reducing the number of required lasers. Micromotion information is characterized by the detuning of cooling lasers and imaging the trajectories of the ions. The results were validated through a molecular dynamics (MD) simulation based on an electric field surface that was constructed by machine learning from the data of COMSOL simulations of the actual experimental conditions. In the following sections, we will detail the experimental setup, discuss methods to control and characterize the motion of the ions, and present the results along with theoretical simulations and discussions.

**EXPERIMENTAL DETAILS**

The experimental setup is illustrated in FIG. 1, featuring a linear quadrupole ion trap (LQT) and an ion fluorescence imaging system. A similar apparatus has been described in detail elsewhere [32,41-43], so a brief description is provided here to introduce the relative directions of the cooling laser, imaging system, and micromotion. Radial confinement is formed by four central electrodes, with two diagonally opposite electrodes (marked as 2 and 3) subjected to both direct current (DC) and RF voltages with a peak-to-peak amplitude of $U_{rf}$ = 324 V$_{pp}$ and a frequency of $\Omega/2\pi$ = 3 MHz, while the remaining two diagonally opposite electrodes (marked as 1 and 4) are subjected to only DC voltages. Axial confinement is provided by the eight segments at both ends with DC voltages ($U_{end}$). The radius of the ion trap is $r_0$= 6.85 mm, and the radius of each electrode is R = 4.51 mm. The trap is enclosed in a UHV chamber of ($<$ $8 \times 10^{-11}$ Torr, Edwards IG40 EX). Be$^+$ ions are generated by laser ablation of metallic beryllium using a focused nanosecond pulsed laser operating at about 2 mJ at 1064 nm (Continuum Minilite II) and subsequently trapped in the linear Paul trap. The trapped Be$^+$ ions are then laser-cooled on the $^2P_{3/2}$ ← $^2S_{1/2}$ (F =2) transition with a linearly polarized 313 nm laser (TOPTICA TA-FHG pro). To prevent decay to the $^2S_{1/2}$ (F=1) state, a repump laser beam, detuned by -1.250 GHz from the cooling laser via acousto-optic modulators (AOMs), is applied to repump population back into the Doppler cooling cycle. The cooling lasers are in the X-Z plane at a 45-degree angle to the X-axis, which is perpendicular to the Y-axis.

The imaging system consists of an electron-multiplying CCD camera (Ando Ultra 888), along with an 8-fold magnification objective lens positioned parallel to the Y-Z plane. The resolution of the camera is 13×13 μm/pixel, resulting in an actual image resolution of 1.625 μm/pixel. The exposure time of the camera is maintained at 300 ms in this experiment. Because the imaging system is perpendicular to the X-axis, and the cooling laser is perpendicular to the Y-axis, micromotion in the X-direction cannot be detected by the imaging system, but it can influence cooling laser detuning. Conversely, micromotion in the Y-direction is detectable by the imaging system but unaffected by the cooling laser detuning, as illustrated in FIG.1B.

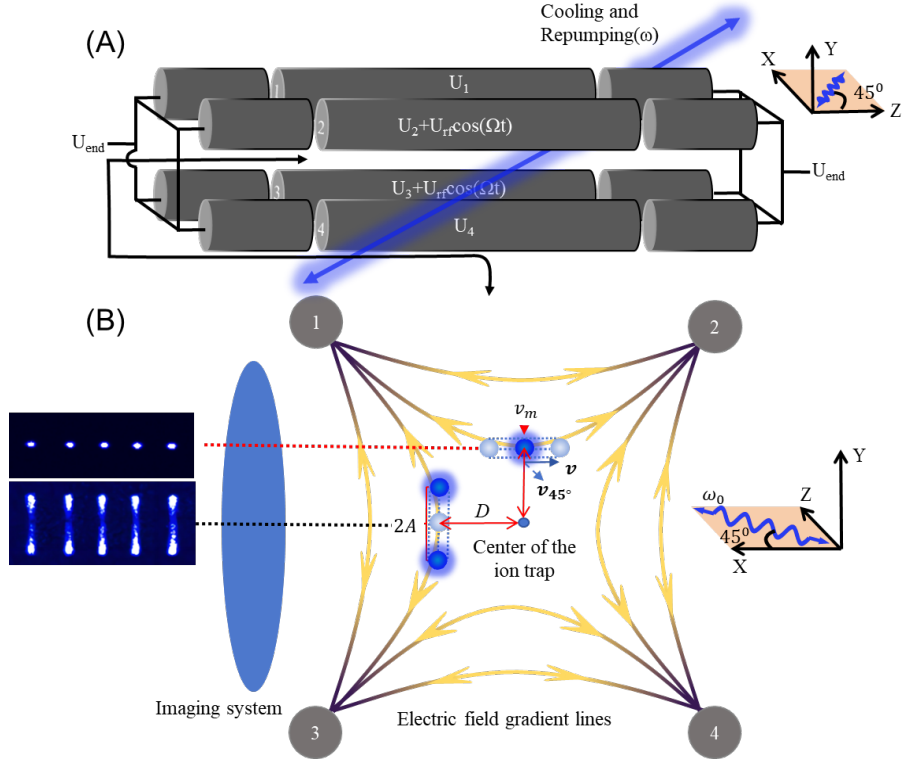

FIG. 1. Schematic of the experimental setup. (A) Voltage configuration on the LQT electrodes: DC is applied to four central electrodes ($U_i$) and all the eight endcaps ($U_{end}$). RF and DC voltages are added to electrodes 2 and 3 with the expression $U_i + U_{rf}\cos(\Omega t)$, where $i$ corresponds to 1, 2, 3, or 4. The cooling lasers are positioned in the X-Z plane at a 45-degree angle to the X-axis, which is perpendicular to the Y-axis. (B) Side view of the LQT: The imaging system is oriented parallel to the Y-Z plane and perpendicular to the X-axis. Micromotion in the Y-direction is measurable by the imaging system, but not in the X-direction.

## THEORETICAL METHOD

The trapped Be$^+$ ions experience multiple forces inside the linear quadrupole Paul trap, and the forces are expressed as follows [44]:

$$\boldsymbol{F}_{\text{Total}} = \boldsymbol{F}_{\text{Trap}} + \boldsymbol{F}_{\text{Static}} + \boldsymbol{F}_{\text{Cooling}} + \boldsymbol{F}_{\text{Coulomb}} \tag{1}$$

where $F_{Trap}$ is the trapping force, $F_{Static}$ is the constant electrostatic force, the laser cooling force is modeled by a damping force $F_{Cooling} = -\alpha \cdot v$ with $\alpha$ as the damping coefficient and $v$ as the ion velocity, and the Coulomb force $F_{Coulomb}$ is the repulsive force from the other ions in the trap. The Coulomb force on the $i$-th ion has the following form,

$$F^i_{Coulomb} = \sum_{j=1, j \neq i}^{N} \frac{Q_i Q_j}{4\pi\varepsilon_0 r_{ij}^2} \hat{r}_{ij} \quad (2)$$

where $Q_i$ and $Q_j$ are the ion charges, $r_{ij}$ is the distance between ions $i$ and $j$, $\hat{r}_{ij}$ is the unit vector pointing from ion $j$ to ion $i$, and $\varepsilon_0$ is the vacuum permittivity.

The motion of the trapped Be$^+$ ions were simulated by MD, which has been proved to be a highly effective tool for studying ion Coulomb crystals [44-49], ion trajectories [50-52] and reaction mechanisms [53,54]. By solving classical equations of motion with the forces in Eq. 1, the trajectories of Be$^+$ ions as well as their velocity and temperatures can be calculated [47,49,55-57].

While the last three forces in Eq. 1 are straightforward to consider, an accurate modeling of the trapping force $F_{Trap}$ plays a crucial role for the description of the motion of the Be$^+$ ions and thus poses some challenges. Traditionally, analytical expression can be used [57,49],

$$E(X, Y, Z, t) = \frac{k_r}{r_0^2} U_{rf} \cos(\Omega t)(x\hat{e}_X - y\hat{e}_Y) + \frac{k_z}{z_0^2} U_{end}(2z\hat{e}_Z - x\hat{e}_X - y\hat{e}_Y) \quad (3)$$

where $r_0$ and $z_0$ are the characteristic dimensions of the trap along the radial ($\hat{e}_X, \hat{e}_Y$) and axial ($\hat{e}_Z$) directions, respectively. $U_{rf}$ and $U_{end}$ are the amplitudes of the RF and DC voltages on the poles, respectively. $\Omega$ is the radio frequency. Normally, additional correction parameters ($k_r$ and $k_z$) are used to consider the structure of the actual Paul trap and the voltages on the trap [49].

In order to accurately model the trapping force in the actual experiment, we here employed a new strategy, namely constructing a time-dependent electric field surface $E(X,Y,Z,t)$ by training a neural network model (shown in FIG. S1) with the dataset of discrete electric field values inside the linear Paul trap. The training dataset were obtained by accurate COMSOL simulations, which accurately considered the geometry of the electrode rods and endcap, their relative positions, and the exerted electric potentials of the actual linear Paul trap.

**RESULTS AND DISCUSSIONS**

**A. Ion Displacement in the X-Direction ($D_X$)**

In the following experiments, we first loaded and laser-cooled approximately 5 $Be^+$ ions at the nodal line of the trap, defined as position 0. We then moved the ions along the X-direction by applying DC voltages ($U_i$) to specified electrodes. For example, by applying higher voltages to electrodes 2 and 4, the ions were pushed closer (positive) to the camera. Note that the cooling lasers are perpendicular to the Y-axis, and the imaging system is parallel to the Y-Z plane. We observed that the ions appeared dumbbell-shaped in the camera (FIG. 2A). As the ions move away from the nodal line of the ion trap along the X-axis, no significant detuning shift in the cooling laser frequency is observed, as shown in FIG. 2B. This indicates that ion motional heating direction is mainly in the Y-axis, as the cooling laser is perpendicular to the Y-direction. The recorded ion motional length (FIG. 2A) increases with X-displacement ($D_X$), exhibiting a linear relationship at a rate of 0.22 μm /μm, as illustrated in FIG. 2D (black dot and linear fit).

Here, the ion motional velocity is calculated from the length (2A) of the oscillating trajectory (shown in FIG 2A). The position of the ions Y(*t*) is: $Y(t) = A\sin(\Omega t + \varphi_0)$, which is shown in FIG. 2C. Hence, the velocity of the ions in the Y-direction $v_Y(t)$ is:

$$v_Y(t) = \Omega A\cos(\Omega t + \varphi_0) = v_{mY} \cos(\Omega t + \varphi_0) \tag{4}$$

The maximum Y-velocity of the ion is:

$$v_{mY} = \Omega A = \beta_X D_X \tag{5}$$

The data is shown as red squares in FIG. 2D. By employing Eq. 5, a linear fit of the experimental data gives $\beta_X = 2.19$ (m/s)/μm, which means the maximum Y-velocity increases up to 1474 m/s with the X-displacement of the ions at a rate of 2.19 (m/s)/μm. When the cooling laser is perpendicular to the motional heating direction, although the velocity of the ions changes due to the micromotion, the population of the excited state remains unchanged. This setup can be particularly useful in the study of state-dependent collisions or temperature-controlled chemical reactions.

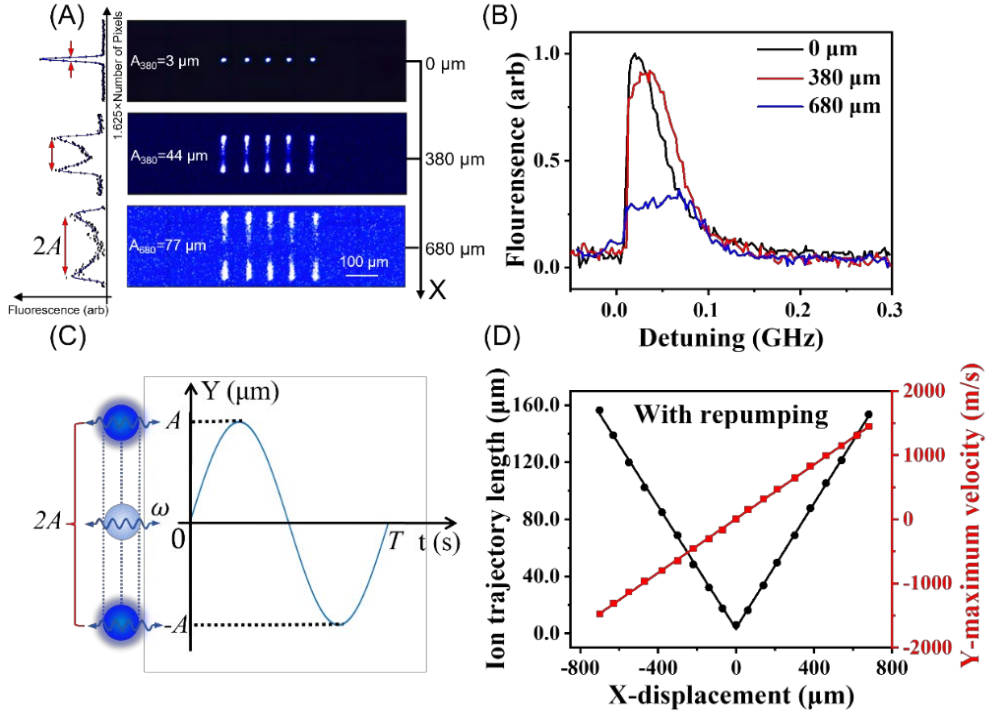

FIG. 2. Displacement of ions in the X-direction ($D_X$) and motional heating in the Y-direction. Larger displacement results in longer motional length of the ions (greater motional heating). (A) Typical ion fluorescence images at different X-displacements. The motional length of ions is determined by calculating the number of pixels corresponding to the ion fluorescence. (B) Cooling laser detuning at different X-displacements. The detuning of the cooling laser frequency increases slightly when the ions are displaced in the X-direction from 0 to 680 μm. (C) Schematic of ion motional position as a function of time, $Y(t) = A\sin(\Omega t + \varphi_0)$. (D) Cooling with repumping. Black dots: The length of the ion trajectory (2A) as a function of X-displacement. A linear fit yields 0.22 μm/μm. Red squares: Maximum velocity ($v_{mY} = \Omega A$) as a function of displacement in the X-direction ($D_X$). The linear fit, $v_{mY} = \beta_X D_X$, shows $\beta_X = 2.19$ (m/s)/μm.

**B. Ion Displacement ($D_Y$) in the Y-direction**

We repeated the above experimental process and then moved the ions along the Y-direction. Note that the cooling lasers are at a 45-degree angle to the X-axis, and the imaging system is perpendicular to the X-axis. As the ions displaced ($D_Y$) further from the trap nodal line, the ions remained visible as clear bright spots from the camera (FIG. 3A). However, the cooling laser detuning increased from 11 MHz to 2.7 GHz, with the ions moved from -583 μm to 583 μm, at a rate of 47 MHz/μm (FIG. 3B, top). This indicates that the micromotion speed of the ions gradually increases mainly in the X-direction.

When the cooling laser detuning exceeds 625 MHz ($D_Y \geqslant 133$ μm), the micromotion-induced Doppler shift (± 625 MHz) can span the two ground-state energy levels (1.25 GHz) of $Be^+$ (FIG. 3C), allowing the ions to be cooled without repumping (FIG. 3A bottom, 3B bottom). The maximum displacement, detuning, and velocity in the trap are constrained by the size and position of the cooling laser spot. By adjusting the cooling laser position as the ions move, we can achieve up to 7.1 GHz laser detuning, which can cover energy levels up to 14.2 GHz without repumping (FIG. S6).

Here, the ion micromotion velocity is determined from the cooling laser detuning. The micromotion, driven by the RF, induces a Doppler shift that modulates the ions' fluorescence rate ($\kappa$), which can be expressed as [22]:

$$\kappa = \kappa_0 \frac{\left(\frac{\Gamma}{2}\right)^2}{\left(\frac{\Gamma}{2}\right)^2 + [\omega_0 - \omega - \omega v_m \cos(\Omega t + \varphi_0)/c]^2} \qquad (6)$$

Where $\Gamma/2\pi$ = 19.4 MHz is the natural linewidth of $Be^+$ $^2P_{3/2}$ state, $\Omega/2\pi$ = 3 MHz is the trap frequency, $v_m$ represents the maximum velocity of the ions in the direction of the cooling laser, $\varphi_0$ is the phase determined by the initial position of the ions relative to the driving field, c is the speed of light, $\omega_0$ is resonance frequency of cooling, $\omega$ is the laser output frequency, and $\delta = \omega_0 - \omega$ refers to the cooling laser detuning. The fluorescence intensity of ions obtained by the camera (*Fluo*) is:

$$Fluo = n\alpha \int_0^T \kappa dt \qquad (7)$$

Where $n$ denotes the number of the ions, and $\alpha$ is the total efficiency of imaging system, determined by the solid angle covered by the reentrant objective, the quantum efficiency of the camera at 313 nm, the exposure time, and the camera's gain. The fluorescence intensity of the ions captured by the camera is strongest when the velocity of the ion is maximum ($\delta = \frac{\omega v_m}{c}$, as shown in FIG. 3A). The calculated results from Eq. 7 are shown in FIG. 3A (middle). The theoretical peaks overlap with the experimental fluorescence peaks by shifting only about 5 MHz, validating that the maximum velocity corresponds to the highest fluorescence of the ions.

Note that the direction of the ion micromotion is at an angle of $\theta$=45° to the cooling laser frequency in this experiment. Hence:

$$v_{mX} = \sqrt{2} \cdot \frac{\delta c}{\omega} = \beta_Y D_Y \qquad (8)$$

Where $v_{mX}$ is the ions' maximum velocity in X-direction. The measured data are shown as the red squares in FIG. 3B, top. The linear fit of the experimental data gives $\beta_Y = 2.09$, which means the ion maximum X-velocity increases with the Y-displacement at

a rate of 2.09 (m/s)/μm. The velocity of the ions at any times, $v_X(t)$, can then be written as:

$$v_X(t) = v_{mX} \cos(\Omega t + \varphi_0) = \beta_Y D_Y \cos(\Omega t + \varphi_0) \qquad (9)$$

The fitted $\beta_Y$ is close to $\beta_X$, indicating that the trap electrical field is symmetric. When the displacements in X and Y directions are the same, the corresponding velocities are similar, and the micromotion heating is primarily in one direction. When the direction of motional heating is perpendicular to the direction of the cooling laser, the detuning remains constant, necessitating the use of repumping lasers in the first case.

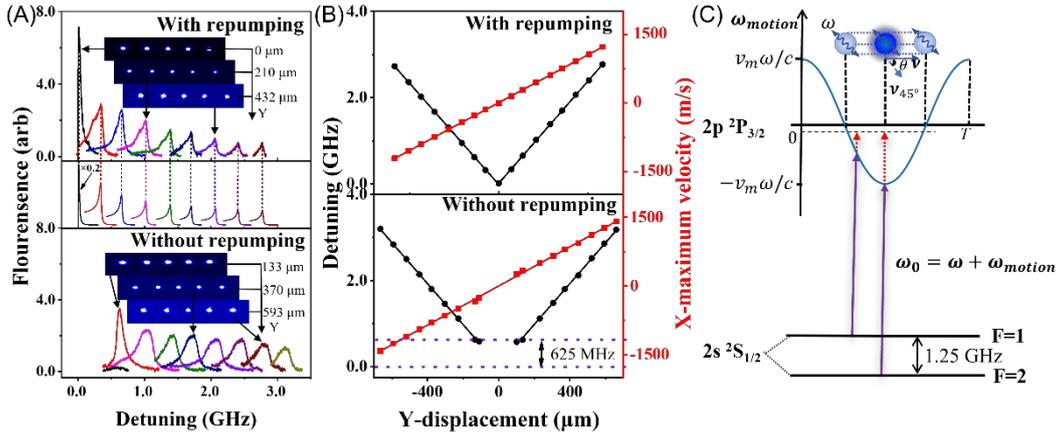

FIG. 3. Displacement of ions in the Y-direction ($D_Y$) and motional heating in the X-direction. Larger displacement results in higher cooling laser detuning (greater motional heating). The trap nodal line is defined as position 0. (A) Typical ions' fluorescence images at different Y-displacements and the corresponding cooling laser detuning. Top: Cooling with repumping. Inserted images show ions at the trap nodal line (0 μm), positive 210 μm, and 432 μm in the Y-direction, corresponding to 0.01 GHz, 1.18 GHz, and 2.07 GHz, respectively. Middle: Numerical solution of Eq. 7 at different values of detuning and maximum speeds. Theoretical peaks are all shifted by about -5 MHz to overlap with experimental results, and the rightmost data is multiplied by a factor of 0.2. Inserted images show ions at 133 μm, 370 μm, and 593 μm in the positive Y-direction, corresponding to 0.63 GHz, 1.80 GHz, and 2.86 GHz, respectively. (B) Relationship among Y-displacement, cooling laser detuning, and maximum velocity in X-direction. Top: Cooling with repumping. Black dots: Cooling laser detuning ($\delta$) as a function of Y-displacement ($D_Y$). A linear fit yields 4.7 MHz/μm. Red squares: Maximum velocity ($v_{mX}$) as a function of displacement ($D_Y$). The linear fit, $v_{mX} = \beta_Y D_Y$, shows $\beta_Y = 2.09$ (m/s)/μm. Bottom: Cooling without repumping. Black dots: Cooling laser detuning ($\delta$) as a function of Y-displacement ($D_Y$). A linear fit yields 4.7 MHz/μm. Red squares: Maximum velocity ($v_{mX}$) as a function of displacement ($D_Y$). The linear fit, $v_{mX} = \beta'_Y D_Y$, shows $\beta'_Y = 2.14$ (m/s)/μm. (C). Energy level diagram of $Be^+$. When sufficient micromotion-induced Doppler shift ($\omega_{motion}$) spans the two ground-state energy levels of $Be^+$, cooling without repumping is achieved.

## C. THEORETICAL RESULTS

MD simulations were performed to simulate the ion motion on the time-dependent electric field surface, *E(X, Y, Z, t)*. First, the accuracy of constructed electric field surface is shown in FIG. S2 of the Supplemental Material. The absolute error of the constructed electric field surface is below 1.5 V/m, indicating the accuracy and capability of the adopted neural network method in reproducing accurate model of the electric field surface.

In the MD simulation, the displacement of the ions was implemented by a constant electrostatic field, which exerts an additional force ($\boldsymbol{F}_{static}$) on the ions and pushes them away from the nodal line of the trap. FIG. 4A shows the electric field vectors on the 2D X-Y plane. The micromotion of the ions is driven by the trapping force $\boldsymbol{F}_{\text{Trap}} = q\boldsymbol{E}(X,Y,Z,t)$. It is straightforward to visualize an additional force on the ions when they are displaced away from the nodal line. The ions then undergo large amplitude micromotion. Specifically, when the ions are displaced along the X-axis, they oscillate harmonically in the perpendicular Y-direction with a frequency of the trapping RF field. Similarly, when the ions are displaced along the Y-axis, they oscillate harmonically in the perpendicular X-direction. Interestingly, when the ions are displaced along the diagonal directions (aligning 45 degrees to the X or Y axis), they oscillate in the same parallel direction. Furthermore, as shown in FIG. 4B and 4C, when the ions are displaced further away from the nodal line, the ions experience a larger electric field and thus undergo a larger amplitude of oscillation and a larger velocity change.

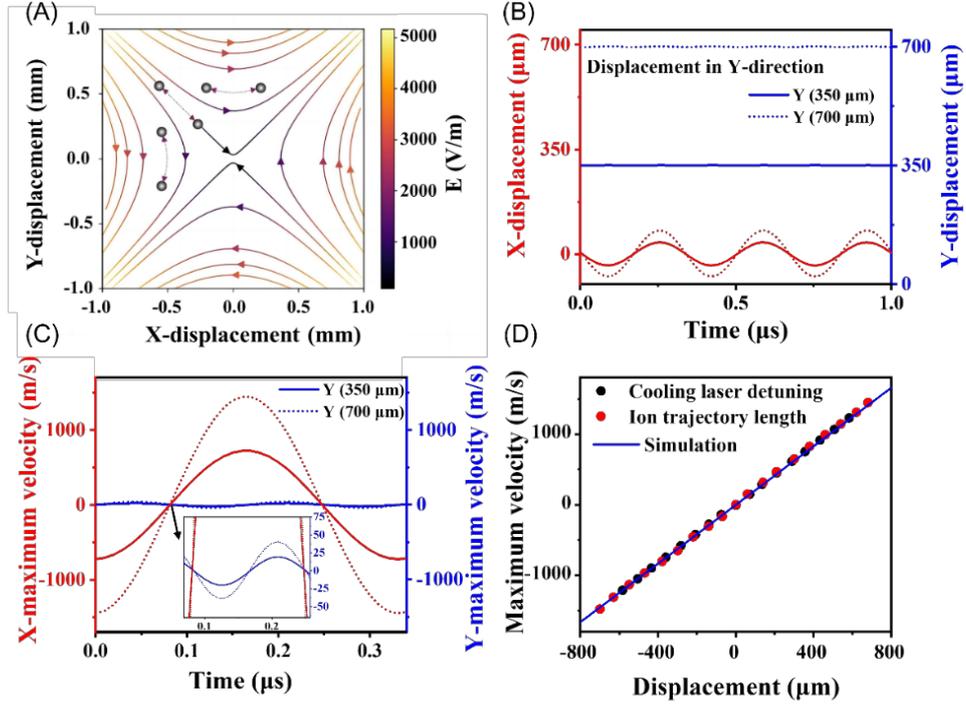

FIG. 4. Theoretical simulation of motional heating through displacement and the comparison with experimental results. (A) Electric field vectors at an arbitrary time and the micromotion of Be$^+$ ions on the 2D X-Y plane cut at Z=0. The electric field is obtained by the neural network model. Trajectories of Be$^+$ ions at different positions are denoted by dashed lines and the Be$^+$ ions are represented as black spheres. (B) The relationship between the coordinates and time when Be$^+$ ions displaced away from the nodal line by distances of 350 μm and 700 μm. (C) Theoretical analysis of the velocity of the Be$^+$ ions over one oscillation cycle when they were displaced away from the nodal line by distances of 350 μm and 700 μm. (D) Comparison of theoretical and experimental results of the maximum velocities at different displacements. Black dots and red dots denote the ion velocity obtained by experimental measurements of cooling laser detuning ($\beta_Y = 2.09$ (m/s)/μm) and the length of the ion trajectory ($\beta_X = 2.19$ (m/s)/μm), respectively. Blue line denotes the simulated ion maximum velocity ($v_{mYs}$)

with respect to the displacement (D$_Y$) and yields a rate of $\beta_{YS} = 2.07$ (m/s)/μm. The comparison of theoretical and experimental results shows a good agreement.

It is interesting to note in the enlarged view in FIG. 4C that the ions also experience a slight heating in the Y-direction when they are displaced in the Y direction. It should also be noted that the oscillation frequency in the Y direction is twice of the RF trapping field. FIG. S3 and S4 in the supporting information show a more detailed trajectories of the ions. In both simulation results, the trajectory of the ion is an arc instead of a straight line, and the height of the arc is about 1.4% of the width. This is too small to be observed in the ion images from the experiment (FIG. 3A), but evidence can be found in the experimental results in FIG. 2B, where a slight shift in cooling laser detuning is observed when the ions are displaced in the X direction, indicating slight heating in the X direction.

To quantitatively compare with experimental observations, the maximum velocity ($v_{mXs}$) of ion motion at different Y-displacements (D$_Y$) were calculated, as shown in FIG. 4D. The simulated results (blue line) give a rate of $\beta_{YS} = 2.07$ (m/s)/μm, which agrees well with the experimental measurements from cooling laser detuning ($\beta_Y = 2.09$ (m/s)/μm) and imaging of ion trajectories ($\beta_X = 2.19$ (m/s)/μm), respectively. This good agreement confirms the equivalence of the two experimental measurements.

**CONCLUSION**

In this study, we achieved laser cooling of Be$^+$ ions without the need for repumping by utilizing one-dimension motional heating. The results were analyzed using cooling laser detuning and imaging of the ion trajectories, both of which yielded consistent results. We also developed a new theoretical simulation method based on machine learning to accurately determine the actual electric field information of the trap. The simulation results were consistent with the experimental findings, providing a reliable theoretical method for exploring the dynamic behavior of ions in traps.

Our findings indicate that when the micromotion vector aligns with the cooling laser axis, the Doppler shift effectively broadens the cooling laser frequency. When the shifting covers the repumping frequency, single-laser cooling of $Be^+$ ions is accomplished. This is particularly beneficial for cooling molecular systems where direct laser cooling is challenging due to the intricate energy structures. Additionally, the micromotion heating is mainly in one direction, allowing for precision measurements experiment in the orthogonal direction of micromotion. Overall, this work not only provides a robust method for managing ion micromotion velocity but also offers insights into laser cooling of complex systems that typically require multiple repumping lasers. Furthermore, it offers a method for controlling collision energy in state-dependent ion-molecule reaction investigations.

**ACKNOWLEDGMENTS**

This work was supported by the National Natural Science Foundation of China (Grant No. 22173040, 22241301, 22103032, 22173042, and 21973037).

# Supplemental Material

## 1. Structure of neural network

The structure of the neural network model is depicted in FIG. S1. The inputs of the model are time *t* and the three-dimensional coordinates (*X, Y, Z*) specifying the positions within the ion trap. The outputs are the three components of the electric field vector. A total of 128 neurons are included in the single hidden layer. The sigmoid function and the Nadam method were used as the activation function and optimizer, respectively. The Nadam optimization algorithm has the advantages of faster convergence speed, adaptive learning rates, and training stability, making it perform exceptionally well in the fitting tasks. The dataset contains a total of 370,440 sample points, which are evenly distributed in the three-dimensional coordinate space and a complete RF period. 20% of these sample points are randomly selected to serve as the validation set. The neural network model were constructed and trained with an in-house Python package that uses available tools such as PyTorch [1] and scikit-learn [2]. By employing this method, we mapped out the electric field surface across the inner space of the ion trap.

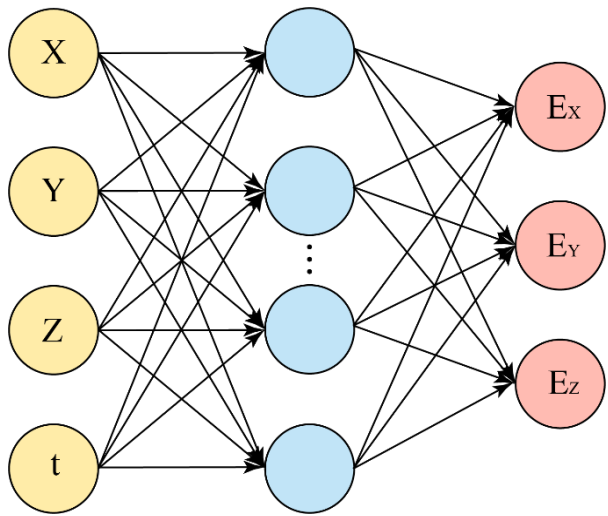

FIG. S1. Structure of the neural network model. The inputs of the model are time $t$ and the three-dimensional coordinates ($X, Y, Z$) specifying the positions within the ion trap. The outputs are the three components of the electric field vector, $E_X$, $E_Y$, $E_Z$. A total of 128 neurons are included in the single hidden layer.

## 2. Analysis of fitting error

To verify the accuracy of our neural network model, we compared the predicted values of the electric field with the original dataset. As shown in FIG. S2, the mean absolute errors ($\Delta E_X$, $\Delta E_Y$) of the electric field intensity along the $X$ and $Y$-directions are below 1.5 V/m, and the mean absolute error ($\Delta E_Z$) along the $Z$ direction is below 0.1 V/m. Importantly, the relative errors in the electric field intensity along the X and Y-directions remain under 0.7% and the RMSE (Root Mean Square Error) is 0.41 V/m. The predictions of our neural network model show a good agreement with the original data, which indicates the accuracy of the adopted neural network.

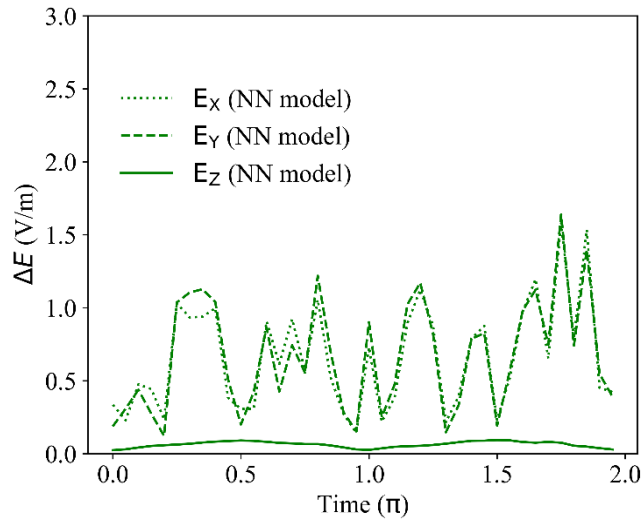

FIG. S2. Absolute error analysis within one RF period ($2\pi$) of the electric field surface predicted by the neural network model, using the original data generated by COMSOL

simulation as references. The dotted line, dashed line, and solid line represent the electric field intensity in the X, Y and Z directions, respectively.

## 3. Additional results of MD simulation

The MD simulation of the ion trajectories was performed with our in-house Python package. It utilized a pre-trained neural network model to predict real-time electric field values at different positions within the ion trap, which are then used to calculate ion trajectories and store information about the ion's trajectory. With an additional superimposed constant uniform electrostatic field, $Be^+$ ions were displaced away from the nodal line of the trap. As shown in Fig. S3, when electric field strengths of 9.75 V/m, 68.51 V/m, and 136.87 V/m were applied in the X-direction, the ion was displaced by 50 μm, 350 μm, and 700 μm, respectively. The ratio of the oscillation amplitudes in the X and Y-directions is about 1.4%.

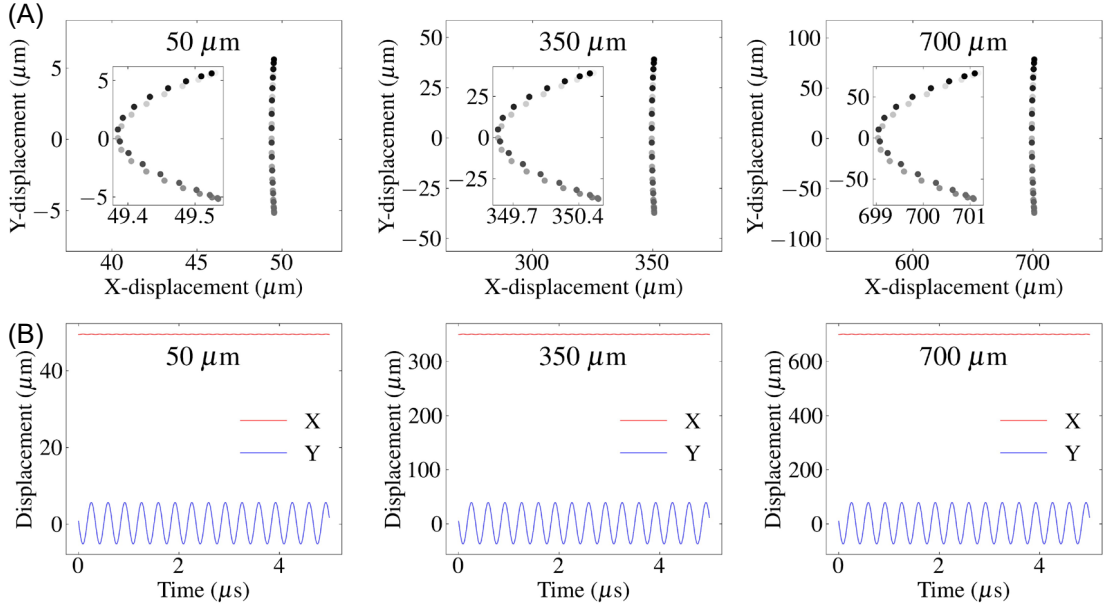

FIG. S3. Ion's motion state when the ion is displaced from nodal line of the ion trap in the X-direction (50 μm, 350 μm, 700 μm) in MD simulations. (A) Trajectories of ion's

motion in the X-Y plane when the ion is displaced from the nodal line by 50 μm, 350 μm, and 700 μm in the X-direction. The range of ion's motion in the Y-direction is 10.82 μm, 76.58 μm, and 153.15 μm, respectively. The motion range in the X-direction is 0.15 μm, 1.07 μm, and 2.13 μm, respectively. The colors from black to light represent different time points of ions within a radio frequency cycle. (B) Displacement of the ion in the X-direction leads to harmonic motion of the ion in the Y-direction with an oscillation frequency equal to the radio frequency.

Similar results were observed when the electric field intensity was applied in the Y-direction, as shown in FIG. S4. The slight differences observed in the results between the two directions were due to the small noise generated during the training process of the neural network model, which resulted in a displaced trap center away from the coordinate origin. This further caused the asymmetric motion of the ion trajectory, as shown for the 50 μm trajectory in FIG. S4.

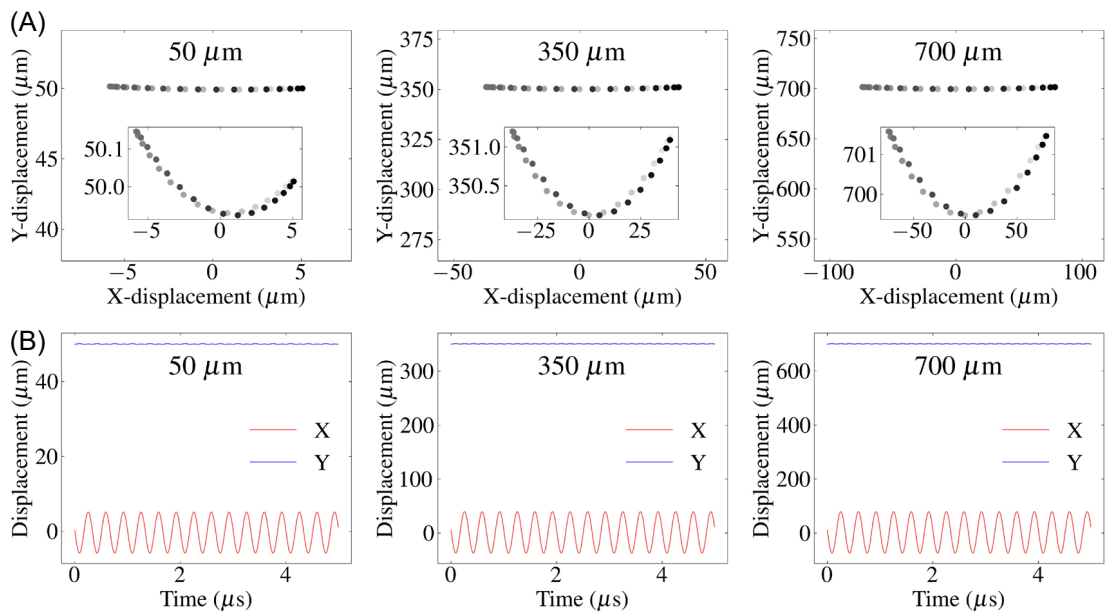

FIG. S4. Ion's motion state when the ion is displaced from the nodal line of the ion trap in the Y-direction (50 μm, 350 μm, 700 μm) in MD simulations. (A) Trajectories of ion's motion in the X-Y plane when the ion is displaced from the nodal line by 50 μm, 350 μm, and 700 μm in the Y-direction. The motion amplitude in the X-direction is 10.92 μm, 76.69 μm, and 153.21 μm, respectively. The motion range in the Y-direction is 0.22 μm, 1.08 μm, and 2.10 μm, respectively. The colors from black to light represent different time points of ions within a radio frequency cycle. (B) Displacement of the ion in the Y-direction leads to harmonic motion of the ion in the X-direction with an oscillation frequency equal to the radio frequency.

However, when the direction of the additional uniformly strong electrostatic field is at a 45-degree angle to the X or Y coordinate axis, the ion motion trajectory is a straight line, as shown in FIG. S5. When electric field strengths of 49 V/m, 2325 V/m, and 10610 V/m were applied at a 45-degree angle to the X or Y-axis, the ion was displaced from the nodal line by 50 μm, 350 μm, and 700 μm, respectively.

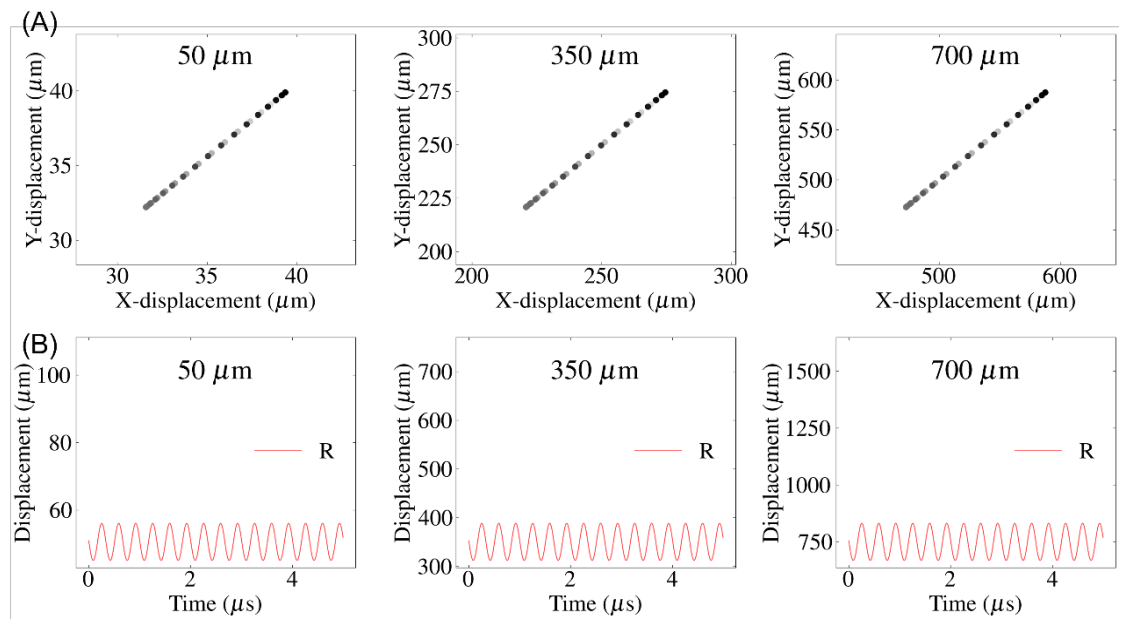

FIG. S5. Ion's motion state when the ion is displaced from the nodal line of the ion trap at a 45-degree angle to the X or Y axis (50 μm, 350 μm, 700 μm) in MD simulations. (A) Trajectories of ion's motion in the X-Y plane when the ion is displaced from the nodal line by 50 μm, 350 μm, and 700 μm at a 45-degree angle to the X or Y axis. The colors from black to light represent different time points of ions within a radio frequency cycle. (B) The distance (R) of the ion from the nodal line as a function of time, and the ion's oscillation frequency are still equal to the radio frequency.

**4. Upper limit displacement of ions without repumping**

To ascertain the limit of ion displacement from the nodal line when ions remain effectively cooled by the laser in the absence of repumping light. We first loaded and laser-cooled several $Be^+$ ions at the nodal line of the trap. Subsequently, we moved the ions along the Y-axis by applying direct current (DC) voltages to the designated electrodes. In this process, the position of the cooling light spot is fine-tuned to ensure that the ions continue to remain within the cooling light spot (without repumping light). In our experiment, the limit of ion displacement from the nodal line is 7.1 GHz ($D_Y$ = 1457 μm), with a corresponding maximum velocity of 3144 m/s (shown in FIG. S6). This can cover a 14.2 GHz energy gap without repumping. If the displacement continues to increase, the background scattering starts to dominate in our system, preventing data collection. To overcome this limitation, a new ion trap system with greater flexibility to move the ion, cooling laser, and imaging system is needed.

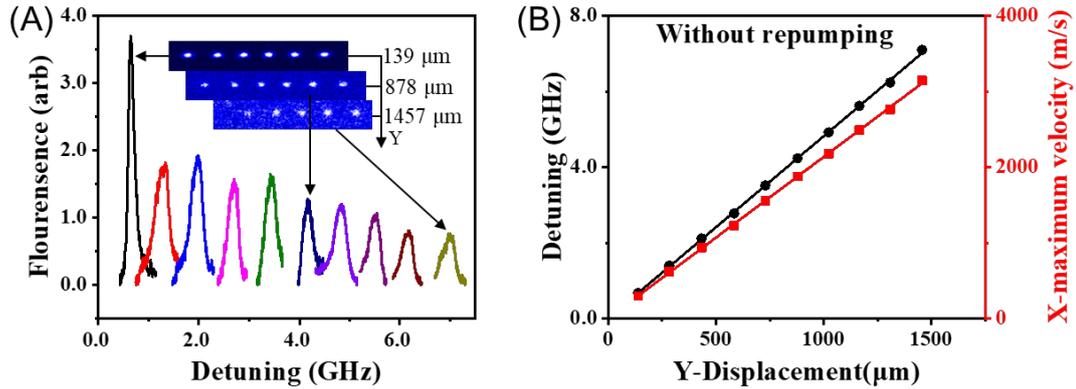

FIG. S6. Upper limit displacement of ions in the Y-direction ($D_Y$) and motional heating in the X-direction without repumping. (A) Typical ions' fluorescence images at large ranges of Y-displacements and the corresponding cooling laser detuning when cooling without repumping. Inserted images show ions at 139 μm, 878 μm, and 1457 μm in the Y-direction, corresponding to 0.68 GHz, 4.24 GHz, and 7.10 GHz, respectively. (B) Cooling without repumping. Black dots: Detuning of the cooling laser when ions are pushed away from the trap center in the Y-direction. Linear fit of the displacement and detuning yields 4.8 MHz/μm. Red squares: calculated ions' maximum velocity and displacement by cooling laser detuning, the linear fit shows 2.13 (m/s)/um.